\documentclass[preprint]{aastex}

\shorttitle{}
\shortauthors{Sembach et al.}
\slugcomment{{\it Accepted for publication in ApJ (June 2002)}}
\begin{document}

\newcommand{\kms}{\,km\,s$^{-1}$}     

\title{A Limit on the Metallicity of Compact High Velocity Clouds }

\author{Kenneth~R.~Sembach\altaffilmark{1}, 
        Brad K. Gibson\altaffilmark{2}, 
 	Yeshe Fenner\altaffilmark{2},
	Mary E. Putman\altaffilmark{3,4}}
\altaffiltext{1}{Space Telescope Science Institute, 3700 San Martin Dr.,
Baltimore, MD  21218}
\altaffiltext{2}{Center for Astrophysics and Supercomputing, Swinburne
University of Technology, P.O. Box 218, Hawthorn VIC 3122, Australia}
\altaffiltext{3}{CASA, University of Colorado, Boulder, CO  80309}
\altaffiltext{4}{Hubble Fellow}

\begin{abstract}
There is a fortuitous coincidence in the positions of the quasar Ton\,S210
and the compact \ion{H}{1} high velocity cloud CHVC\,224.0-83.4-197 on the 
sky.  Using {\it Far Ultraviolet Spectroscopic Explorer} observations of the 
metal-line absorption in this cloud and sensitive \ion{H}{1} 21\,cm emission
observations obtained with the multibeam system at Parkes Observatory, we  
determine a  metallicity of (O/H) $<0.46$ solar at a confidence of 3$\sigma$.  
The metallicity of the high velocity gas is consistent with either an
extragalactic or Magellanic Cloud origin, but is not consistent with a 
location inside the Milky Way unless the chemical history of the gas is
considerably different from that of the interstellar medium
in the Galactic disk and halo. Combined with measurements of highly 
ionized species (\ion{C}{3} and \ion{O}{6}) at high velocities, 
this metallicity limit indicates that the cloud has a substantial halo
of ionized gas; there is as much ionized gas 
as neutral gas directly along the Ton\,S210 sight line.  
We suggest several observational tests that 
would improve the metallicity determination substantially and help
to distinguish between possible origins for the high velocity gas.
Additional observations of this sight line would be valuable since 
the number of compact HVCs positioned in front of background 
sources bright enough for high resolution absorption-line studies is 
extremely limited.
\end{abstract}


\section{Introduction}
 
The lack of distance information for high velocity clouds (HVCs) is a 
long-standing problem in astronomy.  The peculiar kinematics of these clouds
preclude the assumption of a simple velocity-distance relation; generally,
HVCs are defined as parcels of gas moving at $\gtrsim60$ \kms\ in excess of 
velocities expected for interstellar 
gas participating in differential Galactic
rotation (or at $\gtrsim100$ \kms\ with respect to the Local Standard of 
Rest -- LSR). In recent years, there has been renewed interest in the 
kinematics
and distances of HVCs, since the possibility that many of these clouds could be
extragalactic entities has been revisited in the context of embedding baryons
in dark matter halos to provide stability against disruption
(Blitz et al. 1999; Kepner et al. 1999). Extended intragroup distributions of
neutral HVCs with column densities sufficient to produce Lyman-limit 
absorption 
features (N(\ion{H}{1}) $\sim 10^{17}$ cm$^{-2}$) are incompatible with the 
frequency of HVC detections  in other groups (Zwann \& Briggs 2000; Charlton, 
Churchill, \& Rigby 2000), but more confined distributions of compact 
\ion{H}{1} HVCs within a few hundred kiloparsecs of galaxies have not yet 
been ruled out. 

A specific class of HVCs observed in \ion{H}{1} 21\,cm emission, 
the compact HVCs, consists of particularly promising candidates for 
clouds located in the nearby environs of the Milky Way.  The compact HVCs 
have angular
sizes of $\sim1\degr$ (by definition), are isolated from other clouds of 
similar velocity, and as an ensemble have an approximate mean (apparent) infall
velocity  that is similar to that of Local Group dwarf galaxies
(Braun \& Burton 1999).\footnotemark\  
\footnotetext{This is a supportive, but {\it insufficient}, condition of the 
extragalactic scenario (Gibson et al. 2001a).  For example, 
Putman et~al. (2002) present
a recent analysis demonstrating that the ensembles of 
compact HVCs and extended 
HVCs, many of which are likely located within the Galaxy,
have similar velocity centroids in the Local Group Standard of Rest.}
Thermal pressure considerations for compact HVCs that exhibit 
core-envelope structures suggest that some of the 
clouds are located at distances of several hundred kiloparsecs
from the Sun (Braun \& Burton 2000; Burton, Braun, \& Chengalur 2001).
However, these arguments do not account for the presence of highly ionized gas,
which may affect the distance estimates.  A significant fraction of the 
gas may be photoionized if the clouds are exposed to the 
extragalactic ionizing radiation field (see, e.g., Sembach et al. 1999
for some ionization models).
Collisionally ionized gas is also associated with many
HVCs, including at least one compact HVC (Sembach et al. 2000).

An independent assessment of the extragalactic nature of the compact
HVCs can be made by determining the metallicities of the clouds through
ultraviolet absorption-line spectroscopy.  This is a daunting task with 
existing instrumentation because the number of AGNs/QSOs 
bright enough to obtain decent high dispersion ultraviolet spectra
is limited ($\lesssim$150 known sources have 
$F_{1200A} \gtrsim 1\times10^{-14}$ 
erg cm$^{-2}$ s$^{-1}$ \AA$^{-1}$), and the compact HVCs cover only a small 
fraction of the sky ($\sim1$\% of the southern sky is covered by those 
cataloged by Putman et al. 2002).  The only existing measurement of the 
metallicity of a compact HVC is based on \ion{Mg}{2} line strengths
for CHVC+125+41--207 in the direction of Mrk\,205 (Bowen \& Blades 1993; 
Bowen, Blades, \& Pettini 1995).  The measurements indicate that the cloud 
probably has (Mg/H) $<$ 0.2 solar (see Gibson et al. 2001a), but this 
estimate does not account for small-scale structure in the \ion{H}{1} 
distribution, incorporation of Mg into dust grains, or the relative 
ionization of Mg and H.
Small-scale \ion{H}{1} structure in the radio beam
 may increase or decrease (Mg/H) depending upon the exact nature of the 
distribution, while the latter 
two processes would tend to increase the derived value of (Mg/H).

We have identified a second excellent candidate for an ultraviolet 
absorption-line study to determine the metallicity of a compact HVC.
The sight line to the bright
quasar Ton\,S210 ($l=224.97\degr, b=-83.16\degr$, V = 14.7)\footnotemark\
passes through a compact HVC with a velocity $V_{LSR} \approx 
-200$ \kms\ (see Figure~1).  
\footnotetext{Also known as QSO\,0119-286 and GD\,1339 
(Hewitt \& Burbidge 1993), and as X-ray point source 
1RXS\,J012151.5--282048 (Voges et al. 1999).}  
For comparison, Ton\,S210 is a 
factor of 6--7 times brighter at ultraviolet wavelengths than Mrk\,205.
In this paper, we report new \ion{H}{1} 21\,cm measurements of this
cloud, CHVC\,224.0--83.4--197, and place a limit on its metallicity using
extant data from the {\it Far Ultraviolet Spectroscopic Explorer} ($FUSE$).  
 
\section{\ion{H}{1} 21\,cm Observations}

CHVC\,224.0--83.4--197 was originally detected by the \ion{H}{1} Parkes All-Sky
Survey (HIPASS) and classified as a compact HVC (CHVC) in the high velocity
cloud catalog compiled by Putman et al. (2002).  The HIPASS map of this
cloud is shown in Figure~1.  
The CHVC criteria require a diameter
of less than $2\degr$ and no elongation in the direction of nearby extended
emission in the 25\%-of-peak column density contour.
The HIPASS HVC survey has a spatial resolution of $\sim15$\arcmin, a FWHM
velocity
resolution of 26.4 \kms\ (after Hanning smoothing), and a $5\sigma$ \ion{H}{1}
column density sensitivity of $\sim 2.2 \times 10^{18}$ cm$^{-2}$ 
(see Barnes et al. 2001 for a description
of HIPASS).

A series of channel maps constructed from the HIPASS data is shown in 
Figure~2.  The compact HVC toward Ton\,S210
shows up clearly near --200 \kms\ in these data; the cloud was not cataloged
previously by Braun \& Burton (1999) as a compact HVC because its \ion{H}{1} 
column density is slightly lower than the sensitivity limit of their survey.
Two additional compact HVCs are visible in the channel maps.  
CHVC\,197-181.0--184 is located near CHVC\,224.0--83.4--197 and is visible in
the --198 \kms\ and --172 \kms\ channel maps.
The other cloud, CHVC\,221.0--88.2--256 (originally cataloged as 
CHVC\,220--88--265 by Braun \& Burton (1999)), is prominent in the --251 \kms\
and --277 \kms\ channel maps.
Additional noteworthy emission in this region 
of the sky is produced by the Magellanic Stream at $V_{LSR} \approx -100$
\kms\ (Putman 2000). The compact HVC toward 
Ton\,S210 sits $\sim10\degr$ off the Magellanic Stream concentration 
MS~II (Mathewson, Cleary, \& Murray 1975), which corresponds to a projected 
distance of about 9 kpc for a Stream distance of 50 kpc.  
In this region, the Stream has a velocity dispersion of $\approx30$ \kms, so
some of the nearby Stream emission in the maps may extend out to  $V_{LSR}
\approx-150$ \kms.  The velocity of CHVC\,224.0--83.4--197, 
$V_{LSR} \approx -197$ \kms\ (with a width of $\pm20$ \kms), is offset $\sim2\sigma$ from 
the fringes of the Stream emission and $\sim100$ \kms\ from the bulk velocity
centroid of MS~II.
This region of the sky also encompasses
the southern half of the Sculptor 
Group, which contains many dwarf galaxies at $+100 \lesssim V_{LSR} \lesssim 
+600$ \kms\ (C\^ot\'e et al. 1997).

To improve upon the \ion{H}{1} column density measurement directly toward
Ton\,S210, we obtained a high spectral resolution observation of the 
21\,cm emission on 27~April~2001 using the
narrow-band (8\,MHz) filterbank  on the central feed (14.4\arcmin\ FWHP) of
the multibeam system at the Parkes 64\,m telescope.  We used both left- and
right-hand polarizations in the analysis.  The 8~MHz bandwidth spanned
2048 channels, each 0.8~\kms\ wide.  We employed standard
position-switching techniques, with a total of 60 minutes on-source and 60
minutes off-source.  The off-source time was split equally between
two positions with angular throws of $\Delta_\alpha = 10^m$ in right ascension
(east, west).  Sub-exposures were 10 minutes in duration.
To maximize baseline stability, we chose on-source/off-source 
pairs to have identical altitude-azimuth tracks across the sky.

We divided each on-source spectrum  by its companion off-source
spectrum.  The two polarizations were treated separately before combining
the quotient spectra. After fitting and removing a second-order baseline, 
we Hanning-smoothed the final spectrum to 1.6 \kms\ resolution.  The 
root-mean-squared deviation of the resulting spectrum was $\approx7$\,mK
per resolution element.
Velocities were corrected for the Sun's motion relative to the Local
Standard of Rest (a +2.5\,\kms\ transformation), and the native Jy/beam
flux units were converted to brightness temperature via the appropriate scale
factor (0.80~--~Staveley-Smith 2001, private communication).

\section{FUSE Observations}

The $FUSE$ Team observed Ton\,S210 on 1999~October~21 during the commissioning
phase of the mission and again on 2001~August~10.  The light of the quasar 
was centered in the large aperture (LWRS, $30\arcsec\times30\arcsec$) of the 
LiF1 channel during both observations, but the first observation suffered
from poor channel alignment.  We consider only the August 2001 observation 
here.
The observation consisted of 15 exposures totaling 41 ksec of on-source
exposure time.
Approximately 43\% of the data ($\sim17.5$ ksec) was obtained during orbital 
night.  The quasar was intentionally dithered within the LWRS aperture
in the spectral direction
during the course of the observation to sample different portions of the 
microchannel-plate detector.

We processed the time-tagged photon lists with the 
{\tt CALFUSE} (v2.0.5) software available at the Johns Hopkins University in 
December 2001. 
The software screened the data for valid photon events, removed correlated 
burst events, corrected for
geometrical distortions, spectral motions, satellite orbital motion, and
detector background noise, and applied flux and wavelength calibrations.
The fully reduced data have S/N $\sim$ 8--10 per spectral resolution element
and a 
$1\sigma$ wavelength accuracy of $\sim0.024$\,\AA\ ($\sim7$ \kms) set 
by aligning the low velocity portions of the interstellar \ion{Si}{2}
$\lambda1020.699$ and \ion{Ar}{1} $\lambda1048.220$ lines with the 
low velocity portions of the \ion{H}{1} spectrum described in \S2  
(see Figure~1 of Wakker et al. (2001) for an illustration of another 
Ton\,S210 \ion{H}{1} spectrum at low velocities).

In this study, we concentrate our analysis on the data obtained in the LiF1
channel since it has the highest sensitivity of the four $FUSE$ channels
at wavelengths $\lambda > 1000$\,\AA.  
(Near the \ion{O}{1} line at 1039.230\,\AA, it accounts for $\sim50$\%
of the total effective area of the instrument.)
The LiF1 channel covers the 987--1083\,\AA\
and 1094--1187\,\AA\ spectral intervals at a velocity resolution of 
$\sim20$ \kms\ (FWHM).  We use the data from the other three $FUSE$ channels
at these wavelengths to verify that the absorption features observed are 
real and are not detector artifacts.
We also make use of shorter wavelength 
(915--1005\,\AA) data from the SiC2 channel, which provides information
on \ion{H}{1} Lyman series, H$_2$, and \ion{C}{3} absorption.
The 1020--1045\,\AA\ region of the Ton\,S210 spectrum is shown by Savage et al.
(2000).  In Figure~3 we show continuum normalized profiles for several 
interstellar lines observed in the spectrum.  These include \ion{O}{1}
$\lambda1039.230$, \ion{O}{6}$\lambda1031.926$, \ion{C}{2} $\lambda1036.337$,
and \ion{C}{3} $\lambda977.020$.

The $FUSE$ data are archived in the Multi-Mission Archive at the Space 
Telescope Science Institute (MAST) under the dataset name P1070302.  
Information about the $FUSE$ mission and its on-orbit performance can be 
found in papers by Moos et al. (2000, 2002) and Sahnow et al. (2000a, 2000b).

\section{HVC Metallicity}
We derive a metallicity for CHVC\,224.0--83.4--197 by comparing the amount of 
\ion{H}{1} detected in 21\,cm emission with the amount of \ion{O}{1}
observed in absorption.  \ion{O}{1} is an excellent tracer of neutral
gas since the ionization of oxygen and hydrogen are closely coupled 
through strong charge-exchange reactions (Field \& Steigman 1971).  
Therefore, there is no need for an
ionization correction in the metallicity derivation, unlike other HVC 
studies that have relied upon other species (\ion{Mg}{2}, \ion{Si}{2},
\ion{S}{2}, etc.) to estimate metallicities.  Furthermore, oxygen is not 
significantly depleted onto dust grains in low density environments
(Meyer, Jura, \& Cardelli 1998), so the derived value of 
N(\ion{O}{1}) can be compared directly to N(\ion{H}{1}) to derive O/H without 
substantial uncertainties resulting from incorporation of oxygen into solids.

For \ion{H}{1}, we find N(\ion{H}{1}) = 
$(2.2\pm0.2)\times10^{18}$ cm$^{-2}$ by integrating the \ion{H}{1} profile
shown in Figure~3 over the --250 to --150 \kms\ velocity range and 
assuming that the weak emission is optically thin.  
The quoted error on N(\ion{H}{1}) is the statistical noise uncertainty only.
The primary systematic error associated with this determination arises from
the finite beam size of the Parkes emission measurement.  As can be seen in 
Figure~1, the beam may perhaps sample a gradient in the \ion{H}{1} column
within the cloud.  High-resolution imaging of other compact HVCs 
(Braun \& Burton 2000; Burton et al. 2001) 
indicates that some of these clouds contain multiple
cold (T $\sim 100$ K) cores embedded in warmer envelopes.  However, the 
observed width of the \ion{H}{1} line, FWHM $\approx$ 33 \kms, is indicative
of the warmer, turbulent gas expected for the envelopes of the clouds.

As a consistency check on the value of N(\ion{H}{1}) measured from the 
21\,cm emission profile, we examined the $FUSE$ spectrum of Ton\,S210 for 
\ion{H}{1} Lyman series absorption at the velocity of the cloud.  Figure~4 
shows the $FUSE$ LiF1 data for the Ly$\beta$ line and the SiC2 data for 
several higher order lines in the series (Ly$\iota$, Ly$\kappa$, Ly$\lambda$).
We fit the interstellar medium (ISM) absorption with a series of three 
components, 
one based on the low velocity 21\,cm emission [$V_{LSR} \approx -10$ \kms, 
b $\approx 12$ \kms, N(\ion{H}{1}) $\approx 1.5\times10^{20}$ cm$^{-2}$], 
and two broad (b $\approx 20-25$ \kms) sub-Lyman-limit 
[N(\ion{H}{1}) $\sim 5\times10^{16}$ cm$^{-2}$] satellite 
absorbers that are clearly required by the $FUSE$ data but are well below
the detection threshold of the 21\,cm data.  The ISM absorption is represented
by the dashed line in Figure~4.  We added a high velocity component 
[$V_{LSR} = -197$ \kms, b = 25 \kms, N(\ion{H}{1}) = $2.2\times10^{18}$ 
cm$^{-2}$] 
to this model to produce the composite ISM+HVC model represented by the
heavy solid line in Figure~4.  

The $FUSE$ data clearly indicate that high velocity \ion{H}{1} absorption
is present in this direction at a level consistent with that determined from
the 21\,cm emission profile shown in Figure~2.  The $FUSE$ data do not 
allow us to put a tighter constraint on the \ion{H}{1} column density because
we do not know the exact details of the \ion{H}{1} velocity structure in the 
Lyman series lines.  However, for single-component representations of the 
HVC velocity structure, b-values of 20--30 \kms\ and N(\ion{H}{1}) $\gtrsim
1\times10^{18}$ cm$^{-2}$ work best.  Larger b-values start to deviate 
significantly from the 21\,cm line width (b $\approx20$ \kms) and 
overestimate the amount of absorption in the negative velocity wing of the 
observed profile, even for small \ion{H}{1} columns.  

The strongest limit on N(\ion{O}{1}) in our $FUSE$ dataset is set by 
considering the lack of absorption observed in the \ion{O}{1} 
$\lambda1039.230$ line, which has a line strength log\,$f\lambda=0.980$ 
(Wiese, Fuhr, \& Deters 1996).  At the velocities of the HVC, the \ion{O}{1}
line is blended with ISM absorption in the H$_2$ (5--0) R2 line at 
1038.69\,\AA. 
Discounting the H$_2$ absorption and simply integrating over the --230 to
--110 \kms\ velocity range, we find $W_\lambda < 48$~m\AA\, which translates 
to N(\ion{O}{1}) $< 5.5\times10^{14}$ cm$^{-2}$ if {\it all} of the observed
absorption is due to \ion{O}{1} and none is due to H$_2$.
Geocoronal \ion{O}{1} emission does not affect this result since most of the 
observation was obtained at night, and the velocities of the airglow
($V \sim$ 0 \kms) are much smaller than the velocity of the HVC (see Figure~3).

If we model the H$_2$ 1038.69\,\AA\ 
absorption by examining the profiles of other $J=2$ 
lines in the $FUSE$ spectrum (i.e., R2 $\lambda\lambda965.79, 1003.98,
1009.02, 1014.97$; P2 $\lambda\lambda1005.39, 1028.10$; Q2 $\lambda1010.94$),
we find that the 1038.69\,\AA\ line has the shape shown by the solid 
line overplotted on the \ion{O}{1} spectrum in Figure~3.  Removing this
line and integrating the residual absorption yields 
$W_\lambda = 19\pm16$~m\AA\, a result consistent with no detectable \ion{O}{1}
absorption.  If we take the $3\sigma$ uncertainty in this equivalent width
as an upper limit, 48\,m\AA, we again conclude that 
N(\ion{O}{1}) $< 5.5\times10^{14}$ cm$^{-2}$.
This limit accounts for both statistical noise uncertainties and 
continuum placement uncertainties (Sembach \& Savage 1992).

Combining the \ion{O}{1} and \ion{H}{1} results, we obtain 
(O/H) $\le 2.5\times10^{-4}$ (3$\sigma$).  This 
value is $\le0.34$ solar if an oxygen reference abundance of (O/H)$_\odot =
7.41\times10^{-4}$ is used (Grevesse \& Noels 1993) or $\le0.46$ 
solar if (O/H)$_\odot = 5.45\times10^{-4}$ is used (Holweger 2001).  

Now one could also consider whether there are any other metal lines in the 
$FUSE$ bandpass that might be used to put a stronger constraint on the 
metallicity of the gas.  The strongest neutral gas tracer in the $FUSE$
wavelength region is \ion{C}{2} $\lambda1036.337$, which is shown in 
Figure~3.  
For the observed \ion{H}{1} column,
we expect a \ion{C}{2} $\lambda1036.337$ line strength of $\sim110$ m\AA\
assuming a b-value of 10 \kms, a solar ratio of (C/H), 
(C/H)$_\odot = 3.89\times10^{-4}$ (Holweger 2001), and no more than 
a factor of 
2 depletion of carbon onto dust grains (Sofia et al. 1997; Sofia \& 
Meyer 2001). 
High velocity \ion{C}{2} is present along the sight line at a level of 
$W_\lambda \approx 180$\,m\AA, but the strongest absorption occurs at 
velocities about 25 \kms\ less than those of the peak \ion{H}{1} 
21\,cm emission.
This velocity offset is significantly larger than the expected uncertainty in
the $FUSE$ wavelength calibration, so we consider several possible 
explanations for this offset.

First, the \ion{C}{2} absorption may be associated with neutral gas seen at 
this velocity in the channel maps shown in Figure~2.  The presence of 
\ion{C}{2} in this component combined with a reduction in line strength near 
--197 \kms, where the \ion{H}{1} 21\,cm emission peak occurs, suggests that 
the \ion{C}{2} may be tracing a more diffuse component of the neutral gas.
There is some velocity substructure within the 21\,cm profile consistent
with such an interpretation.  The velocity differences in the profiles 
might be related to substructure within the 
HVC on scales smaller than the radio telescope beam size.  Depending on how
the \ion{C}{2} and \ion{H}{1} profiles are related to each other, the 
metallicity derived from this comparison (assuming all of the \ion{C}{2}
arises in {\it neutral} gas) could be anywhere in the range from $<0.1$ solar
up to solar.

Second, the \ion{C}{2} absorption may really be a redshifted intergalactic 
medium absorber along the $0.116\pm0.001$ redshift path to the quasar 
(Wisotski et al 2000).  One possibility is a Ly$\beta$ absorber 
at $z=0.00976$, which could eventually be checked by searching for the 
corresponding Ly$\alpha$ absorption at 1227.55\,\AA\ with longer wavelength
data from the {\it Hubble Space Telescope} ($HST$). 
(The corresponding Ly$\gamma$ line, if present, is blended with 
the interstellar H$_2$ (0--0) R1 line at 982.07\,\AA.)  Other possibilities
are Ly$\gamma$ absorption at 
$z=0.06498$ or \ion{C}{3} at $z=0.0610$, but neither have corresponding 
Ly$\beta$ lines in the SiC2 spectra at 1092.38\,\AA\ or 1087.36\,\AA,
respectively. High order Lyman series absorptions at 
higher redshifts can be excluded by noting that the corresponding 
Ly$\beta$ absorption is not present in the $FUSE$ spectrum.

Finally, the option we consider most likely is that 
the \ion{C}{2} may be primarily tracing 
ionized gas in the vicinity of the HVC.  Higher ionization gas traced by 
\ion{C}{3} and \ion{O}{6} is present at velocities similar to those of the 
\ion{C}{2} (Figure~3). In this case, there may be little \ion{H}{1}
or \ion{O}{1} directly associated with the \ion{C}{2}.  The 
cloud could have a low metallicity as inferred from the absence of \ion{O}{1}
in the neutral portions of the cloud
and still exhibit \ion{C}{2} absorption if a substantial amount (e.g., 50\%)
of the gas associated with the cloud is ionized (see \S5).

We conclude that the observed feature at --175 \kms\ in the \ion{C}{2} 
spectrum is in fact \ion{C}{2} associated with CHVC\,224.0--83.4--197,
but that it traces predominantly ionized gas rather than neutral gas.
Therefore, this absorption does not place a strong 
constraint on the metallicity of the cloud but may prove useful in later
studies of the ionization properties of the high velocity gas.  We adopt a 
metallicity (O/H) $\le 0.46$ solar (3$\sigma$) for 
CHVC\,224.0--83.4--197.

\section{Discussion}

The metallicities of only a few HVCs have been measured precisely.  These 
include the Magellanic Stream, which has $Z\sim0.2-0.3$ (Lu et al. 
1994; Sembach et al. 2001; Gibson et al. 2000) and 
Complex~C, which has $Z \sim0.1-0.2$ (Wakker et al. 1999;
Richter et al. 2001) or possibly  as high as  $Z \sim0.3$ in some 
directions (Gibson et al. 2001b).\footnotemark\
\footnotetext{The metallicity $Z$ is given in units of the solar value 
($Z_\odot=1$).}
In both HVCs,
this high velocity gas is
located well beyond the confines of the Galactic disk. 
A distant halo location is possible for Complex~C, but the Magellanic 
Stream is clearly outside the Galaxy.   
Even though at this time we are  only able to impose a 3$\sigma$ 
upper limit of $Z \lesssim0.46$ on the metallicity of CHVC\,224.0--83.4--197, 
this bound is consistent with an external location for this compact HVC as 
well.  It is also consistent with a Magellanic origin,
but the velocity does not match the bulk Stream velocity well in this 
direction. 

Our limit on (O/H) in CHVC\,224.0--83.4--197 is $\sim2\sigma$ 
lower than the value (O/H)$_{LISM} = 
(3.19\pm0.37)\times10^{-4}$ found by Meyer et al. 
(1998) for the local interstellar medium\footnotemark.  
\footnotetext{The error on this quantity is the standard deviation of the 
weighted mean, not the smaller error in the weighted mean (0.14) quoted by the 
authors.}
It is unlikely that differential oxygen depletion onto dust
grains is responsible for the 
lower value of  O/H seen in the HVC compared to the  
nearby gas since the interstellar value of O/H appears to be very constant 
over a range of environmental parameters (Meyer et al. 1998; Cartledge et al.
2001).  Although the metallicity limit for CHVC\,224.0--83.4--197 
suggests that there is a difference between O/H measured locally and O/H 
in the HVC, it is not yet possible to state unambiguously that the 
HVC is extragalactic in nature.  
However,  it is possible that the metallicity
could in fact be considerably lower than the limit implies.  This hypothesis 
is easily testable --
a much more rigorous limit on the metallicity of CHVC\,224.0--83.4--197 
could be 
obtained directly by measuring the \ion{O}{1} $\lambda1302.168$ line
with the $HST$ since the line strength, $f\lambda$, 
is a factor of 6.7 larger than that of the 1039.230\,\AA\ line observed by
$FUSE$.
 
One caveat in determining the metallicity of the HVC is the possibility that 
the \ion{H}{1} emission traced by our 21\,cm observations is not representative
of the \ion{H}{1} content directly along the infinitesimal beam subtended 
directly along the sight line by the absorption-line measurements.  
If the HVC \ion{H}{1} is cold and very clumpy, 
the Ton\,S210 sight line might pass near the emitting clumps 
without penetrating them.  In such a scenario, it would be difficult to 
explain the substantial \ion{H}{1} Lyman-series absorption detected at the 
HVC velocities by $FUSE$.
This possibility also seems unlikely given the 
relatively broad \ion{H}{1} 21\,cm profile width, the compact nature of the 
cloud as a whole, and the exponential halo 
distributions of the gas seen in other compact HVCs (Burton et al. 2001).
However, the structure of the \ion{H}{1} 21\,cm emission in the region 
should be checked with deep, high angular resolution imaging.  

The presence of ionized species (\ion{C}{3}, \ion{O}{6}) at velocities 
very near to those of the \ion{H}{1} HVC suggests that the Ton\,S210 sight line
penetrates an outer, ionized layer of the cloud in addition to the neutral
cloud regions seen in the \ion{H}{1} Lyman series absorption.  This type of  
multi-phase ionization structure may be common in HVCs.  
Sembach et al. (2000) have suggested that 
the \ion{O}{6} arises in the boundary between cool / warm \ion{H}{1} gas 
and a hotter (T$\sim10^6$ K) external medium, such as a hot Galactic 
corona or Local Group medium.  A hot, pervasive intergalactic medium is 
predicted by models of the formation of large 
scale structure in the presence of cold dark matter (e.g.,
 Cen \& Ostriker 1999a, 1999b; Dav\'e et al. 1999).  
A rough estimate of the amount of ionized gas associated with the HVC can be 
made by considering the strength of the \ion{O}{6} $\lambda1031.926$
profile shown in Figure~3.  Integrating the optical depth of this line 
from --250 \kms\ to --110 \kms\ yields N(\ion{O}{6}) 
$ = (8.7\pm1.5)\times10^{13}$ cm$^{-2}$.  The corresponding amount of
\ion{H}{2} is given by  
N(H$^+$)~=~N(\ion{O}{6})~(O/H)$_\odot^{-1}~Z^{-1}~f_{OVI}^{-1}$.
For an \ion{O}{6} ionization fraction $f_{O VI} < 0.2$ (see Tripp, Savage,
\& Jenkins 2000), 
N(H$^+$) $> 8\times10^{17} Z^{-1}$. Assuming the metallicity of 
the ionized gas is comparable to that of the neutral gas ($Z < 0.46$),
we find N(H$^+$) $> 1.7\times10^{18}$ cm$^{-2}$, which implies that there is
probably as much \ion{H}{2} as \ion{H}{1} associated with the HVC directly 
along the Ton\,S210 sight line.  The total amount of \ion{H}{2} in the HVC
depends upon the (unknown) distribution of the ionized gas within the cloud,
and may well represent a sizable fraction of the total gas content.

The combination of metallicity determinations and studies of the ionization
properties of HVCs should help to clarify their place in the cosmic web 
of gas that forms galaxies.  
While our observations of the Ton\,S210 sight
line are consistent with an extragalactic location for CHVC\,224.0--83.4--197,
we must await planned observations with the $HST$ and ground-based 
measurements of the H$\alpha$ emission from the ionized gas in the cloud
to determine whether the HVC is very low metallicity gas left over from the 
formation of the Local Group or debris from the tidal interactions of the 
Milky Way and Magellanic Clouds.

\smallskip
We thank the members of the $FUSE$ operations and science teams for their 
dedicated efforts to develop and operate $FUSE$.  We also thank Bill Oegerle,
Todd Tripp, and Bart Wakker for helpful comments.
Partial financial support was provided by NASA contract NAS5-32985
and NASA Long Term Space Astrophysics grant NAG5-3485 (KRS).


\newpage
\clearpage
\begin{figure}[ht!]
\includegraphics{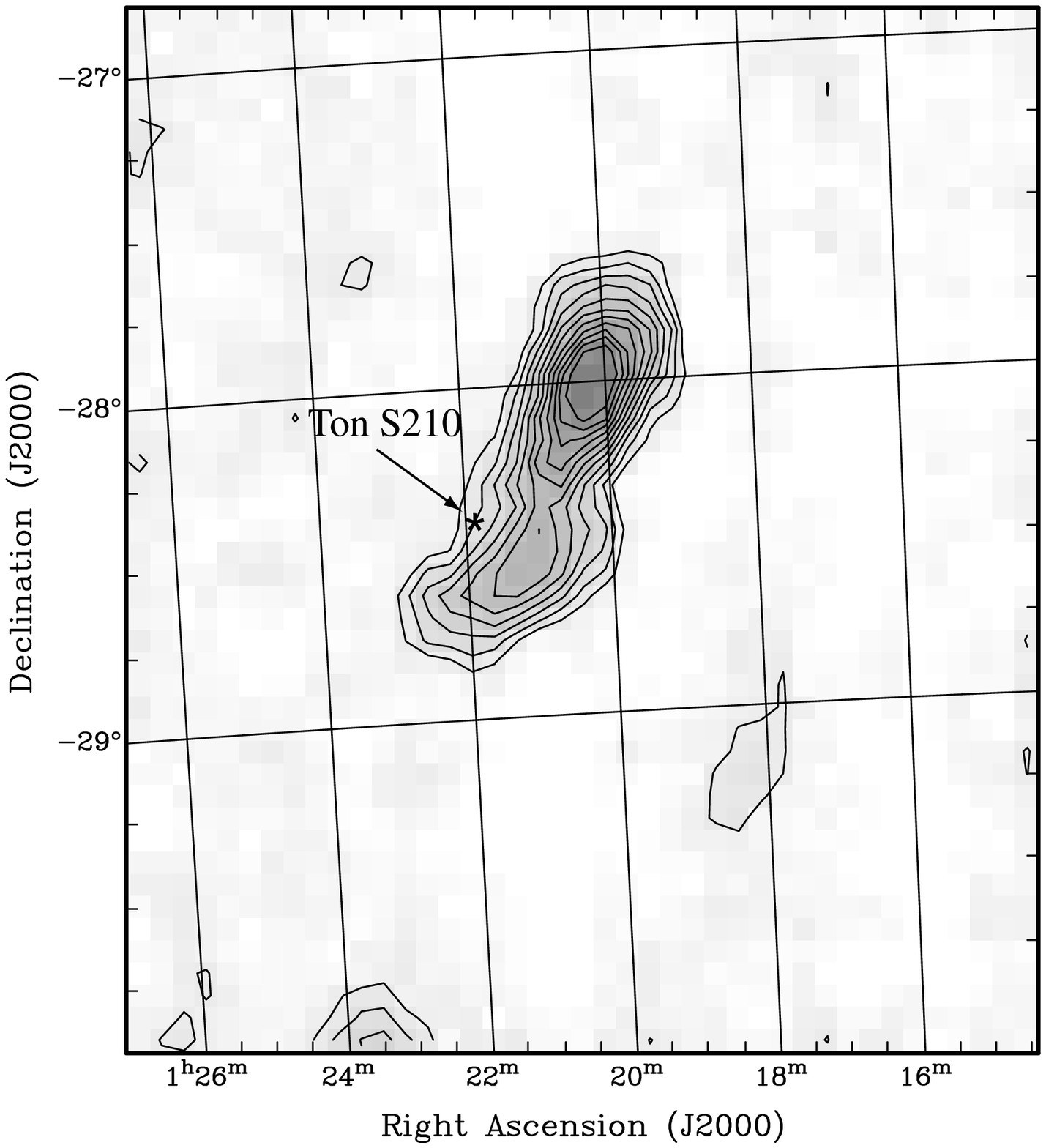}
\vspace{5.8in}
\caption{An \ion{H}{1} 21cm integrated intensity map centered on an LSR 
velocity of 
--200 \kms\ showing the compact high velocity cloud CHVC\,224.0--83.4--197
in the direction of 
Ton\,S210 (Putman et al. 2001).  The contours shown correspond to column
densities of (1.0--11.0)$\times10^{18}$ cm$^{-2}$.
This HVC is considered compact
and isolated under both the original
Braun \& Burton (1999) and refined Putman et~al. (2001) definitions.
It has a total \ion{H}{1} mass of $\sim1.7\times10^4$\,M$_\odot$ at 50 kpc and 
$\sim6.6\times10^6$ M$_\odot$ at 1~Mpc.  This region of the sky 
contains the Magellanic Stream ($V_{LSR} \approx -100$ \kms),
the southern half of the Sculptor Group, and positive and negative velocity
clouds from --250 to +250 \kms\ (Putman 2000).  Ton~S210 lies near the edge of 
the CHVC\,224.0--83.4--197. The HIPASS data used to construct this map are 
fully sampled at an angular resolution of 15.5\arcmin.}
\end{figure}

\newpage
\clearpage
\begin{figure}[ht!]
\vspace{5.2in}
\caption{ {\bf [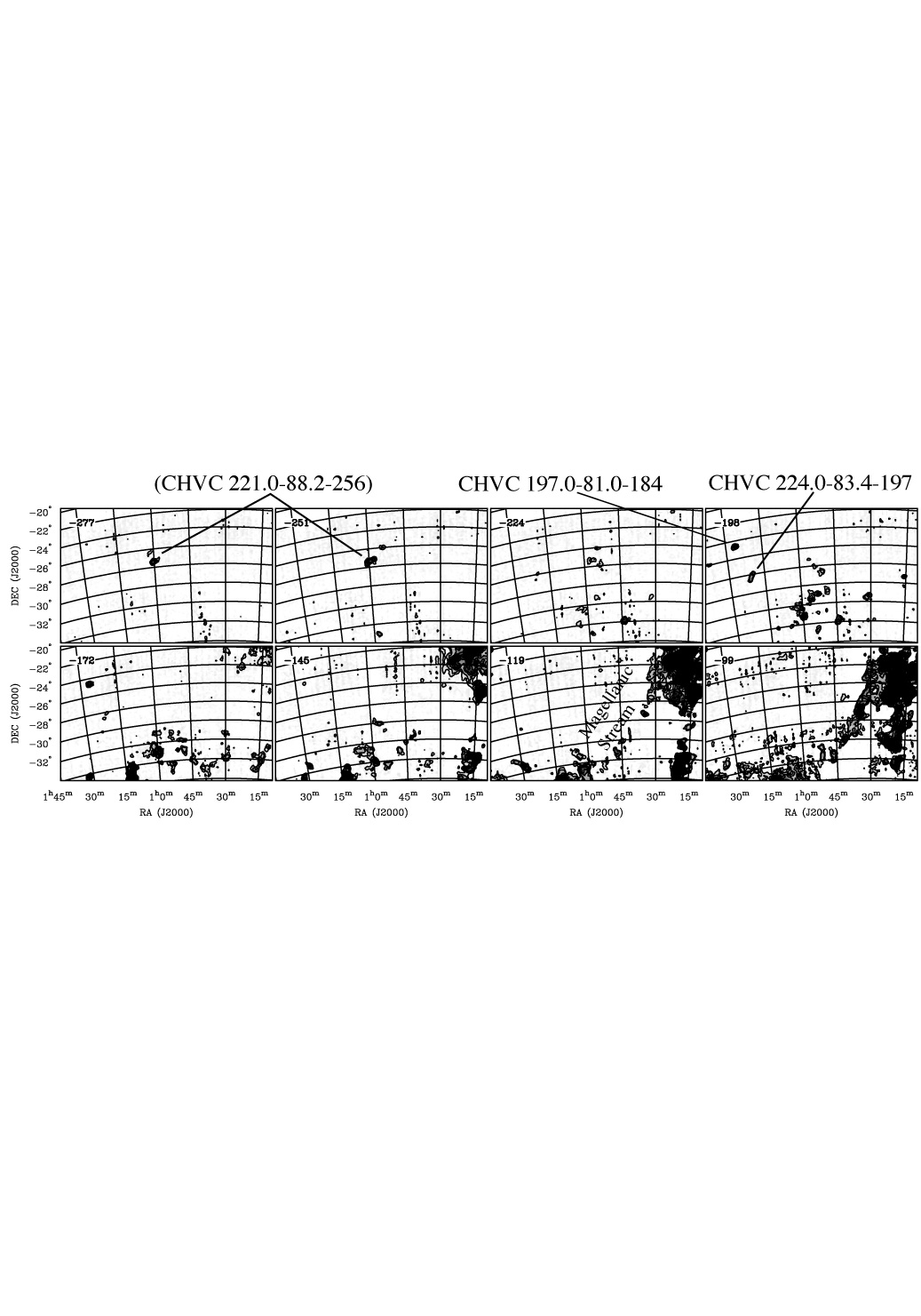]} HIPASS channel maps of the \ion{H}{1} 21cm emission in the region
of sky surrounding Ton\,S210.  The channel velocity (in \kms) is indicated 
in the upper left corner of each panel.  CHVC\,224.0--83.4--197 is visible at 
velocities 
of $\sim -170$ to --225 \kms\ (upper right panel). Two other compact HVCs 
cataloged by Putman et al. (2001), CHVC\,221.0--88.2--256 and 
CHVC 197.0--81.8--184, are visible in several channels. 
The Magellanic Stream has velocities of 
less than $\sim-150$ \kms\ and is prominent near $\sim-100$ \kms\
(lower right panels).
The contours shown correspond to brightness temperatures of 0.02, 0.04,
0.06, etc. Kelvins.}
\end{figure}

\newpage
\clearpage
\begin{figure}[ht!]
\includegraphics{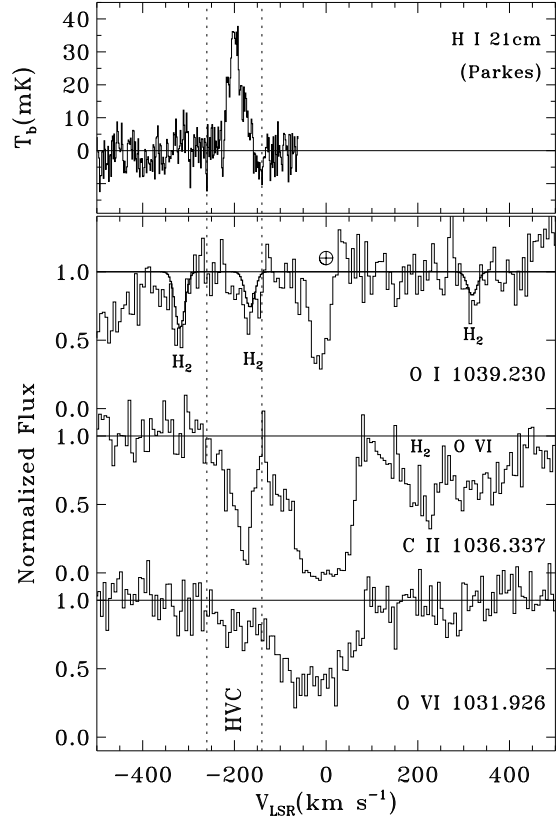}
\vspace{5.5in}
\caption{Continuum normalized interstellar 
absorption-line profiles versus LSR velocity
for selected lines observed toward Ton\,S210 by $FUSE$.
The wavelength of each line is identified near each spectrum; 
additional absorption features from nearby lines (mainly H$_2$) are also 
identified.  Absorption by Milky Way halo gas occurs near 0 \kms, while 
absorption associated with the HVC should occur near --200 \kms.
For the \ion{O}{1} $\lambda1039.230$ spectrum, we have shown both the 
day+night (solid) and night-only (dotted) profiles since the low velocity
portion of the profile is affected by geocoronal \ion{O}{1} airglow emission.
The expected absorption due to the H$_2$ (5--0) P1 $\lambda1038.16$ and 
R2 $\lambda 1038.69$ lines is shown as a smooth dark line on top of the 
\ion{O}{1} spectrum.  The top panel shows the deep, position-switched Parkes
multibeam spectrum obtained for this study.}
\end{figure}

\newpage
\clearpage
\begin{figure}[ht!]
\includegraphics{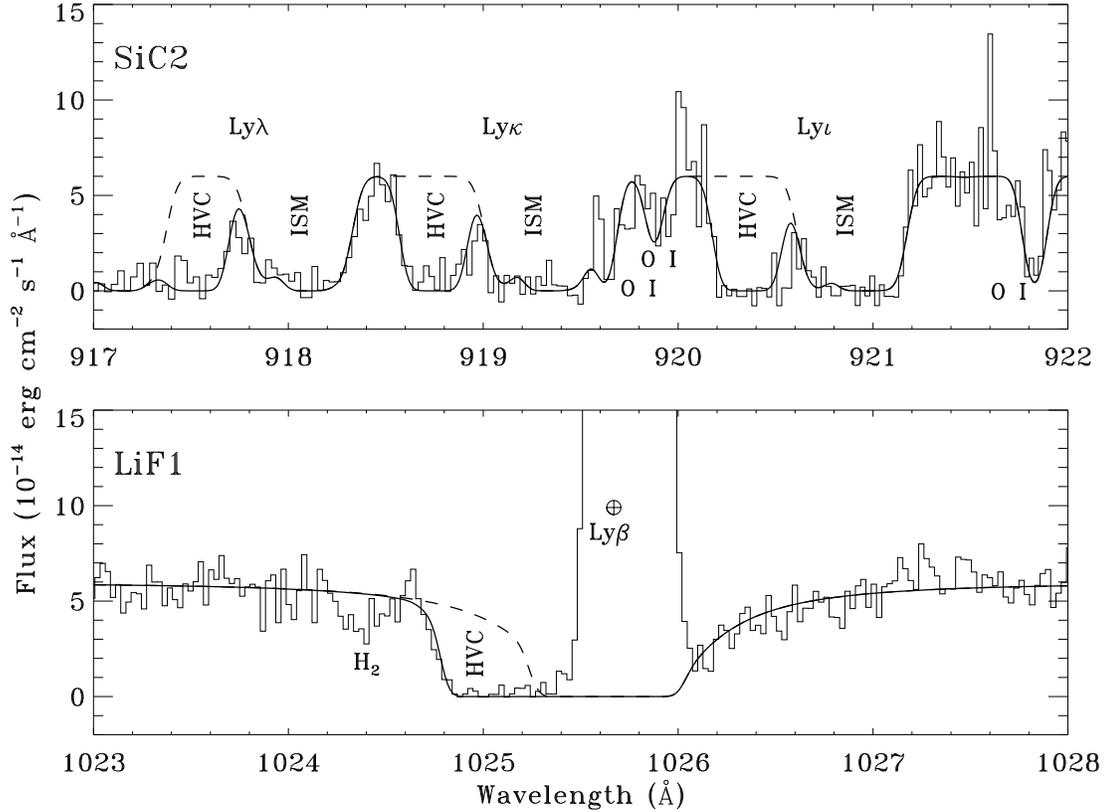}
\vspace{5.5in}
\caption{Selected \ion{H}{1} Lyman series lines observed by $FUSE$ toward 
Ton\,S210.  The dashed line indicates the absorption expected for interstellar
gas in the Milky Way disk and halo (see text).  The heavy solid line indicates
the absorption expected when a high velocity component associated with 
CHVC\,224.0--83.4--197 is included in the model.  The observed Lyman series
absorption features at the velocities of the HVC are consistent with 
the cloud properties inferred from the \ion{H}{1} 21\,cm emission data shown
in Figure~3.}
\end{figure}

\end{document}